\documentclass{article}
\usepackage{amssymb}

\usepackage{amsmath}
\usepackage{doublespace}

\input{tcilatex}

\begin{document}

\title{On the pseudo-Hermitian nondiagonalizable Hamiltonians}
\author{G. Scolarici\thanks{%
e-mail: scolarici@le.infn.it} and L. Solombrino\thanks{%
e-mail: solombrino@le.infn.it} \\
Dipartimento di Fisica dell'Universit\`{a} di Lecce \\
and INFN, Sezione di Lecce, I-73100 Lecce, Italy}
\maketitle

\begin{abstract}
We consider a class of (possibly nondiagonalizable) pseudo-Hermitian
operators with discrete spectrum, showing that in no case (unless they are
diagonalizable and have a real spectrum) they are Hermitian with respect to
a semidefinite inner product, and that the pseudo-Hermiticity property is
equivalent to the existence of an antilinear involutory symmetry. Moreover,
we show that a typical degeneracy of the real eigenvalues (which reduces to
the well known Kramers degeneracy in the Hermitian case) occurs whenever a
fermionic (possibly nondiagonalizable) pseudo-Hermitian Hamiltonian admits
an antilinear symmetry like the time-reversal operator $T$. Some
consequences and applications are briefly discussed.

\end{abstract}

\section{Introduction}

Non Hermitian Hamiltonians play by now a relevant r\^{o}le in physics, in
that they appear in several completely different problems \cite{mie}. Among
them, a remarkable subclass is given by the\textit{\ pseudo-Hermitian
operators }\cite{lee}\textit{, i.e.}, those operators which satisfy 
\begin{equation}
\eta H\eta ^{-1}=H^{\dagger }
\end{equation}%
with $\eta =\eta ^{\dagger }$ [instead, whenever (1) holds without any
constraint on the (linear and invertible) operator $\eta $, $H$ is called 
\textit{weakly pseudo-Hermitian} \cite{so}]. Of course, Hermiticity is a
particular case of pseudo-Hermiticity, corresponding to $\eta =\mathbf{1}.$
Pseudo-Hermiticity also represents the mathematical background of a recent
proposal on a \textit{complex extension of Quantum Mechanics }\cite{be,mv}.

The essential feature of the pseudo-Hermitian operators is the peculiarity
of their spectrum, which can be constituted by real as well as complex (but
grouped in complex-conjugate pairs) eigenvalues \cite{so,m1}. This property,
originally stated with reference to diagonalizable operators with discrete
spectrum, has been recently extended to a class of nondiagonalizable
Hamiltonians \cite{m6}. Such Hamiltonians can arise, for instance, for some
critical parameter values, whenever a physical system undergoes a
perturbation which preserves the pseudo-Hermiticity, but not the
diagonalizability, of its Hamiltonian. An example of such a situation is
shown in Sec. 5 .

The aim of this paper is just to carry on a systematic and deep study on
nondiagonalizable pseudo-Hermitian operators.

To this end, we recall (and partly refine) in Sec. 2 the basic results on
this subject. Next, we inquire in Sec. 3 into the definiteness or the
indefiniteness of the metric induced by $\eta $ , concluding that for any
pseudo-Hermitian operator $H$ with discrete spectrum, the metric is always
indefinite unless $H$ is diagonalizable with real spectrum. This result
disproves a recently stated theorem on the subject \cite{m6}.

Successively, in Sec. 4, we take into account another characteristic feature
of the pseudo-Hermiticity property,\textit{\ i.e.}, its connection with the
existence of antilinear symmetries, which has been already enlightened in
the case of diagonalizable operators \cite{so,m3}, showing that such
connection holds also for the nondiagonalizable case. Sec. 5 is devoted to a
discussion on the time-reversal invariance of fermionic Hamiltonians,
extending a result on the (generalized) Kramers degeneracy that we have
already proven for diagonalizable operators \cite{kr}. Finally, some
concluding remarks and possible applications of the previous results are
briefly presented in Sec. 6.

\section{The spectrum of nondiagonalizable pseudo-Hermitian operators}

According to \cite{m6} we consider here only linear operators $H$ acting in
a separable Hilbert space $\mathfrak{H}$ and having discrete spectrum.
Moreover, throughout this paper we shall assume that all the eigenvalues $%
E_{n}$ of $H$ have finite algebraic multiplicity $g_{n}$ and that there is a
basis of $\mathfrak{H}$ in which $H$ is block-diagonal with
finite-dimensional diagonal blocks. Then, a complete biorthonormal basis $%
\mathfrak{E}=\left\{ \left| \psi _{n},a,i\right\rangle ,\left| \phi
_{n},a,i\right\rangle \right\} $ exists such that the operator $H$ can be
written in the following form \cite{m6}:%
\begin{equation}
H=\sum_{n}\sum_{a=1}^{d_{n}}(E_{n}\sum_{i=1}^{p_{n,a}}\left| \psi
_{n},a,i\right\rangle \left\langle \phi _{n},a,i\right|
+\sum_{i=1}^{p_{n,a}-1}\left| \psi _{n},a,i\right\rangle \left\langle \phi
_{n},a,i+1\right| )
\end{equation}%
where $d_{n}$ denotes the geometric multiplicity (\textit{i.e.}, the degree
of degeneracy) of $E_{n}$, $a$ is a degeneracy label and $p_{n,a}$
represents the dimension of the simple Jordan block $J_{a}\left(
E_{n}\right) $ associated with the labels $n$ and $a$ (hence, $%
\sum_{a=1}^{d_{n}}p_{n,a}=g_{n}$). Furthermore, we denote by $k\left(
n,a\right) $ the total number of identical simple blocks $J_{a}\left(
E_{n}\right) $ occurring in the above decomposition of $H$.

Hence, $\left| \psi _{n},a,1\right\rangle $ (respectively, $\left| \phi
_{n},a,p_{n,a}\right\rangle $) is an eigenvector of $H$ (respectively, $%
H^{\dagger }$ ): 
\begin{equation}
H\left| \psi _{n},a,1\right\rangle =E_{n}\left| \psi _{n,}a,1\right\rangle
,\qquad H^{\dagger }\left| \phi _{n},a,p_{n,a}\right\rangle =E_{n}^{\ast
}\left| \phi _{n,}a,p_{n,a}\right\rangle ,
\end{equation}%
and the following relations hold:%
\begin{eqnarray}
H\left| \psi _{n},a,i\right\rangle &=&E_{n}\left| \psi _{n},a,i\right\rangle
+\left| \psi _{n},a,i-1\right\rangle ,\quad i\neq 1, \\
\qquad H^{\dagger }\left| \phi _{n},a,i\right\rangle &=&E_{n}^{\ast }\left|
\phi _{n},a,i\right\rangle +\left| \phi _{n},a,i+1\right\rangle ,\text{ \ }%
i\neq p_{n,a}.
\end{eqnarray}

The elements of the biorthonormal basis obey the usual relations:

\begin{equation}
\langle \psi _{m},a,i|\phi _{n},b,j\rangle =\delta _{mn}\delta _{ab}\delta
_{ij},
\end{equation}%
\begin{equation}
\sum_{n}\sum_{a=1}^{d_{n}}\sum_{i=1}^{p_{n,a}}\left| \psi
_{n},a,i\right\rangle \left\langle \phi _{n},a,i\right|
=\sum_{n}\sum_{a=1}^{d_{n}}\sum_{i=1}^{p_{n,a}}\left| \phi
_{n},a,i\right\rangle \left\langle \psi _{n},a,i\right| =\mathbf{1.}
\end{equation}

The following theorem has been proven in \cite{m6} :

\textbf{Theorem 1.} \textit{Let }$H$ \textit{be a linear operator acting in
a Hilbert space }$\mathfrak{H}$\textit{. Suppose that the spectrum of }$H$ 
\textit{is discrete, that its eigenvalues have finite algebraic
multiplicity, and that (2) holds. Then, the following conditions are
equivalent:}

\textit{i}) \textit{the eigenvalues of }$H$ \textit{are either real or come
in complex-conjugate pairs and the geometric multiplicity and the Jordan
dimensions of the complex-conjugate eigenvalues coincide;}

\textit{ii}) $H$ \textit{is pseudo-Hermitian.}

In order to fix our notation, and for the benefit of the reader, we prefer
to provide here a (somewhat different) proof of the implication $%
i)\Rightarrow ii)$, which allows us to obtain a useful decomposition of $%
\eta .$

Let us therefore assume that condition $i)$ holds, and use (whenever it is
necessary) the subscript ''$_{0}$'' to denote real eigenvalues, and the
subscripts ''$_{\pm }$'' to denote the complex eigenvalues with positive or
negative imaginary part, respectively. Then, $H$ assumes the following form
(see Eq. (2)):%
\begin{eqnarray}
H
&=&\sum_{n_{0}}\sum_{a=1}^{d_{n_{0}}}(E_{n_{0}}\sum_{i=1}^{p_{n_{0},a}}%
\left| \psi _{n_{0}},a,i\right\rangle \left\langle \phi _{n_{0}},a,i\right|
+\sum_{i=1}^{p_{n_{0},a}-1}\left| \psi _{n_{0}},a,i\right\rangle
\left\langle \phi _{n_{0}},a,i+1\right| )+  \notag \\
&&\sum_{n_{+},n_{-}}\sum_{a=1}^{d_{n_{+}}}[%
\sum_{i=1}^{p_{n_{+},a}}(E_{n_{+}}\left| \psi _{n_{+}},a,i\right\rangle
\left\langle \phi _{n+},a,i\right| +E_{n_{-}}\left| \psi
_{n_{-}},a,i\right\rangle \left\langle \phi _{n_{-}},a,i\right| )+  \notag \\
&&\sum_{i=1}^{p_{n_{+},a}-1}(\left| \psi _{n_{+}},a,i\right\rangle
\left\langle \phi _{n+},a,i+1\right| +\left| \psi _{n_{-}},a,i\right\rangle
\left\langle \phi _{n_{-}},a,i+1\right| )].
\end{eqnarray}

Furthermore, given any complete orthonormal basis $\mathfrak{F}=\{\left|
u_{n},a,i\right\rangle \}$ in our space (that we denote by the same $n,a,i$
labels used for the elements of $\mathfrak{E}$ ), let us pose

\begin{equation}
S=\sum_{n}\sum_{a=1}^{d_{n}}\sum_{i=1}^{p_{n,a}}\left| \phi
_{n},a,i\right\rangle \left\langle u_{n},a,i\right| \text{ },
\end{equation}%
and $\tilde{H}=S^{\dagger }HS^{\dagger \text{ }-1}$. By a straightforward
calculation one obtains

\begin{eqnarray}
\tilde{H} &=&\sum_{n_{0}}\sum_{a=1}^{d_{n_{0}}}(E_{n_{0}}%
\sum_{i=1}^{p_{n_{0},a}}\left| u_{n_{0}},a,i\right\rangle \left\langle
u_{n_{0}},a,i\right| +\sum_{i=1}^{p_{n_{0},a}-1}\left|
u_{n_{0}},a,i\right\rangle \left\langle u_{n_{0}},a,i+1\right| )+  \notag \\
&&\sum_{n_{+},n_{-}}\sum_{a=1}^{d_{n_{+}}}[%
\sum_{i=1}^{p_{n_{+},a}}(E_{n_{-}}\left| u_{n_{+}},a,i\right\rangle
\left\langle u_{n+},a,i\right| +E_{n_{+}}\left| u_{n_{-}},a,i\right\rangle
\left\langle u_{n_{-}},a,i\right| )+  \notag \\
&&\sum_{i=1}^{p_{n_{+},a}-1}(\left| u_{n_{+}},a,i\right\rangle \left\langle
u_{n+},a,i+1\right| +\left| u_{n_{-}},a,i\right\rangle \left\langle
u_{n_{-}},a,i+1\right| )].
\end{eqnarray}

Then, let us consider the involutory operators $U$ and $V$ defined
respectively as follows:

\begin{equation}
U|u_{n_{\pm }},a,i\rangle =|u_{n_{\mp }},a,i\rangle ,\text{ \ \ \ \ \ \ }%
U|u_{n_{0}},a,i\rangle =|u_{n_{0}},a,i\rangle \text{\ ,}
\end{equation}%
and

\begin{equation}
V|u_{n},a,i\rangle =|u_{n},a,p_{n,a}+1-i\rangle .
\end{equation}

The explicit forms of $U$ and $V$ are:

\begin{equation}
U=U\mathbf{1}=\sum_{n_{0},a,i}\left| u_{n_{0}},a,i\right\rangle \left\langle
u_{n_{0}},a,i\right| +\sum_{n_{+},n_{-},a,i}(\left|
u_{n_{-}},a,i\right\rangle \left\langle u_{n_{+}},a,i\right| +\left|
u_{n_{+}},a,i\right\rangle \left\langle u_{n_{-}},a,i\right| )
\end{equation}%
and

\begin{align}
V& =V\mathbf{1}=\sum_{n_{0},a,i}\left|
u_{n_{0}},a,p_{n_{0},a}+1-i\right\rangle \left\langle u_{n_{0}},a,i\right| +
\notag \\
& \sum_{n_{+},n_{-},a,i}(\left| u_{n_{+}},a,p_{n_{+},a}+1-i\right\rangle
\left\langle u_{n_{+}},a,i\right| +\left|
u_{n_{-}},a,p_{n_{-},a}+1-i\right\rangle \left\langle u_{n_{-}},a,i\right| ).
\end{align}%
Moreover both $U$ and $V$ are clearly Hermitian operators, and (recalling
that, by hypothesis, $p_{n_{+},a}=p_{n_{-},a}$)

\begin{eqnarray*}
UV &=&UV\mathbf{1}=\sum_{n_{0}}\sum_{a=1}^{d_{n_{0}}}%
\sum_{i=1}^{p_{n_{0},a}}|u_{n_{0}},a,p_{n_{0},a}+1-i\rangle \left\langle
u_{n_{0}},a,i\right| + \\
&&\sum_{n_{+},n_{-}}\sum_{a=1}^{d_{n_{+}}}\sum_{i=1}^{p_{n_{+},a}}(\left|
u_{n_{+}},a,p_{n_{+},a}+1-i\right\rangle \left\langle u_{n_{-}},a,i\right|
+\left| u_{n_{-}},a,p_{n_{-},a}+1-i\right\rangle \left\langle
u_{n_{+}},a,i\right| )=VU.
\end{eqnarray*}

Thus, one can easily verify that $\tilde{H}$ is a pseudo-Hermitian operator:%
\begin{equation*}
\tilde{\eta}\tilde{H}\tilde{\eta}^{-1}=\tilde{H}^{\dagger }
\end{equation*}%
where $\tilde{\eta}=UV.$ Hence, finally,%
\begin{equation*}
\eta H\eta ^{-1}=H^{\dagger }
\end{equation*}%
where

\begin{eqnarray}
\eta &=&S\tilde{\eta}S^{\dagger }=SUVS^{\dagger
}=\sum_{n_{0}}\sum_{a=1}^{d_{n_{0}}}\sum_{i=1}^{p_{n_{0},a}}|\phi
_{n_{0}},a,p_{n_{0},a}+1-i\rangle \left\langle \phi _{n_{0}},a,i\right| + 
\notag \\
&&\sum_{n_{+},n_{-}}\sum_{a=1}^{d_{n_{+}}}\sum_{i=1}^{p_{n_{+},a}}(\left|
\phi _{n_{+}},a,p_{n_{+},a}+1-i\right\rangle \left\langle \phi
_{n_{-}},a,i\right| +\left| \phi _{n_{-}},a,p_{n_{-},a}+1-i\right\rangle
\left\langle \phi _{n_{+}},a,i\right| )  \notag \\
&=&\eta ^{\dagger }.
\end{eqnarray}

In conclusion we see that \textit{the spectrum of a pseudo-Hermitian
operator is real if and only if }$U\equiv \mathbf{1}$\textit{\ \ (hence, by
Eq. (15), }$\eta =SVS^{\dagger }$)\textit{, }and that \textit{a
pseudo-Hermitian operator is diagonalizable if and only if }$V\equiv \mathbf{%
1}$\textit{\ \ (hence, again by Eq. (15), }$\eta =SUS^{\dagger }$)\textit{.}

\bigskip

\textbf{Remark.}

We stress here that in order to prove the implication $ii)\Longrightarrow i)$
only the invertibility of $\eta $ is needed, while the Hermiticity property $%
\eta =\eta ^{\dagger }$ does not come into play \cite{m6}. Hence, by the
same arguments one can prove that even the spectrum of a \textit{weakly
pseudo-Hermitian} operator \cite{so} [\textit{i.e.}, an operator which
satisfies Eq.(1) without any constraint on the (linear and invertible)
operator $\eta $], satisfies condition $i)$. On the other hand, the above
proof shows that condition $i)$ implies that an Hermitian operator $\eta $
exists which fulfils Eq. (1). Thus, if we just consider operators having a
discrete spectrum, the (possibly broader) class of weakly pseudo-Hermitian
operators actually coincides with the one of pseudo-Hermitian operators.
Nevertheless, we recall that the weak pseudo-Hermiticity is a more useful
notion, in that, for instance, it simplifies checking Eq. (1).

\section{Nondiagonalizability and metric indefiniteness}

We have seen in the previous section that an Hermitian operator $\eta $
always exists such that a nondiagonalizable operator $H$ (whose spectrum
obeys condition $i)$ in Theorem 1) is pseudo-Hermitian; moreover, it is well
known that in this case one can define a new inner product \cite{m6} 
\begin{equation}
\left\langle \left\langle \psi ,\phi \right\rangle \right\rangle _{\eta
}:=\left\langle \psi \right| \eta \left| \phi \right\rangle ,
\end{equation}%
and, correspondingly, a $\eta $-pseudonorm $\left\langle \left\langle \psi
,\psi \right\rangle \right\rangle _{\eta }$. Then, one may of course inquire
into the definiteness or the indefiniteness of the metric induced by $%
\left\langle \left\langle ,\right\rangle \right\rangle _{\eta }.$

Eq. (15) in the previous section clearly shows that the metric associated
with such an $\eta $ cannot be a definite (nor a semidefinite) operator;
indeed, being $\tilde{\eta}$ involutory and non-identical (unless $H$ is a
diagonalizable operator with real spectrum), some of its eigenvalues (but
not all) must be negative, hence the same happens (by the Sylvester's law of
inertia \cite{ho1} ) for the eigenvalues of the operator $\eta $. This fact
can suggest that in all cases of nondiagonalizable (or else, diagonalizable
with complex spectrum) pseudo-Hermitian operators, the metric must be
indefinite; however, as Eq. (15) does not provide us the more general form
of $\eta $ , we must resort to some other argument in order to confirm this
conjecture.

Let us then consider the simplest $2\times 2$ nondiagonalizable operator $A$
: 
\begin{equation*}
A=\left( 
\begin{array}{cc}
E & 1 \\ 
0 & E%
\end{array}%
\right) (E\in \mathbf{R}).
\end{equation*}%
By a straightforward calculation one can verify that $A$ is pseudo-Hermitian
and the \textit{more general} operator $\eta $\ which fulfils Eq. (1) is 
\begin{equation*}
\eta =\left( 
\begin{array}{cc}
0 & k \\ 
k & k^{\prime }%
\end{array}%
\right) (k\neq 0);
\end{equation*}%
moreover $\eta =\eta ^{\dagger }$ if and only if $k,k^{\prime }\in \mathbf{R}
$. The eigenvalues of $\eta $ have with certainty opposite signs, and
obviously the same happens for the $\eta $-pseudonorm of the corresponding
eigenvectors; hence some state exists with a negative $\eta $-pseudonorm,
beside other states with a positive $\eta $-pseudonorm.

This simple example disproves a recently stated theorem according to which
`` (a nondiagonalizable operator) $H$\textit{\ is pseudo-Hermitian if and
only if it is Hermitian with respect to a positive semi-definite inner
product}'' \cite{m6}.

Actually, the following theorem holds.

\textbf{Theorem 2.} \textit{Let }$H$ \textit{be a }$\eta $-\textit{%
pseudo-Hermitian operator with discrete spectrum. Then, the operator }$\eta $%
\textit{\ is definite if and only if }$H$ \textit{is diagonalizable with
real spectrum.}

\textbf{Proof. }Let $H$ be a pseudo-Hermitian operator. We preliminarily
observe that, being in any case $\eta $ an invertible operator, all its
eigenvalues must be different from zero, so that the metric induced by the
inner product (16) either is definite or is indefinite. Now, let us suppose
that a positive (respectively, negative) definite operator $\eta $ exists
which fulfils condition (1); then, an $R$ exists such that $\eta =R^{\dagger
}R$ (respectively, $\eta =-R^{\dagger }R$) \cite{ho1} , and by Eq. (1) we
obtain

\begin{equation*}
RHR^{-1}=R^{\dagger -1}H^{\dagger }R^{\dagger }=(RHR^{-1})^{\dagger },
\end{equation*}%
\textit{i.e.}, $RHR^{-1}$ is Hermitian, hence it is diagonalizable and it
has a real spectrum. Since the similarity transformations preserve the
properties of the spectrum, the same occurs for $H$. Conversely, if $H$ is
diagonalizable with real spectrum, then by Eq. (15) in the previous section
a positive definite metric $\eta =SS^{\dagger }$ exists which fulfils
condition (1) (since in this case $U=V\equiv \mathbf{1}$ ) .$\blacksquare $

\section{Nondiagonalizable operators and antilinear symmetries}

\bigskip

A very intriguing feature of the pseudo-Hermiticity property is its
connection with the existence of antilinear symmetries. This connection was
already acknowledged to hold in the case of diagonalizable operators with
discrete spectrum \cite{so,m3} ; indeed, the pseudo-Hermiticity property is
a necessary and sufficient condition for a (diagonalizable) operator $H$ to
admit an antilinear (involutory) symmetry \cite{so}. Considering the great
physical interest in the study of such symmetries (we recall that the
time-reversal symmetry is associated, in complex quantum mechanics, with an
antilinear operator), we intend here to inquire the above-mentioned
connection in the case of nondiagonalizable pseudo-Hermitian operators. To
this end, let us premise a definition.

\textbf{Definition.}\cite{asc} \textit{Given the complete orthonormal basis }%
$\mathfrak{F}=\{\left| u_{m},a,i\right\rangle \}$ \textit{in a Hilbert
space, we call }conjugation \textit{associated with it the involutory
antilinear operator}

\begin{equation}
\Theta _{\mathfrak{F}}=\sum_{m,a,i}\left| u_{m},a,i\right\rangle K\langle
u_{m},a,i|,
\end{equation}%
\textit{where the operator }$K$\textit{\ acts transforming each complex
number on the right into its complex conjugate.}

Analogously, in the case of a complete biorthonormal basis $\mathfrak{E}%
=\left\{ \left| \psi _{n},a,i\right\rangle ,\left| \phi
_{n},a,i\right\rangle \right\} $, we call conjugation associated with it the
involutory antilinear operator \cite{so}%
\begin{equation}
\Theta _{\mathfrak{E}}=\sum_{n,a,i}\left| \psi _{n},a,i\right\rangle
K\left\langle \phi _{n},a,i\right| .
\end{equation}

Then, the following theorem holds.

\textbf{Theorem 3.} \textit{Let }$H$ \textit{be a linear operator. Suppose
that the spectrum of }$H$ \textit{is discrete, that its eigenvalues have
finite algebraic multiplicity and that (2) holds. Then the following
conditions are equivalent:}

\textit{i) an antilinear invertible operator }$\Omega $ \textit{exists such
that }$[H,\Omega ]=0$\textit{;}

\textit{ii) }$H$\textit{\ is (weakly) pseudo-Hermitian;}

\textit{iii) an antilinear involutory operator }$\widehat{\Omega }$\textit{\
exists such that }$[H,\widehat{\Omega }]=0$\textit{;}

\textit{iv) a basis exists in which }$H$\textit{\ assumes a real form.}

\bigskip

\textbf{Proof.} $i)\Rightarrow ii)$. Let $\Omega $ exist such that $%
[H,\Omega ]=0$. This implies that $[\tilde{H},\tilde{\Omega}]=0$, where $%
\tilde{H}=S^{\dagger }HS^{\dagger \text{ }-1}$ and $\tilde{\Omega}%
=S^{\dagger }\Omega S^{\dagger -1}$. Then, the linear operator

\begin{equation*}
\tilde{\eta}=V\Theta _{\mathfrak{F}}\tilde{\Omega}
\end{equation*}%
(where $\mathfrak{F}$ is the orthonormal basis associated with $\tilde{H}$
(see Eq. (10)), while $V$ and $\Theta _{\mathfrak{F}}$ are defined as in
Eqs.(12) and (17), respectively) fulfils the condition stated by Eq. (1),
hence $\tilde{H}$ is (weakly) pseudo-Hermitian; indeed,

\begin{multline}
V\Theta _{\mathfrak{F}}\tilde{\Omega}\tilde{H}\tilde{\Omega}^{-1}\Theta _{%
\mathfrak{F}}^{-1}V^{-1}=V\Theta _{\mathfrak{F}}\tilde{H}\Theta _{\mathfrak{F%
}}V=  \notag \\
=\sum_{n_{0}}\sum_{a=1}^{d_{n_{0}}}(E_{n_{0}}\sum_{i=1}^{p_{n_{0},a}}\left|
u_{n_{0}},a,i\right\rangle \left\langle u_{n_{0}},a,i\right|
+\sum_{i=1}^{p_{n_{0},a}-1}\left| u_{n_{0}},a,i+1\right\rangle \left\langle
u_{n_{0}},a,i\right| )+  \notag \\
\sum_{n_{+},n_{-}}\sum_{a=1}^{d_{n_{+}}}[%
\sum_{i=1}^{p_{n_{+},a}}(E_{n_{+}}^{\ast }\left| u_{n_{+}},a,i\right\rangle
\left\langle u_{n+},a,i\right| +E_{n_{+}}\left| u_{n_{-}},a,i\right\rangle
\left\langle u_{n_{-}},a,i\right| )+  \notag \\
\sum_{i=1}^{p_{n_{+},a}-1}(\left| u_{n_{+}},a,i+1\right\rangle \left\langle
u_{n_{+}},a,i\right| +\left| u_{n_{-}},a,i+1\right\rangle \left\langle
u_{n_{-}},a,i\right| )]=\tilde{H}^{\dagger },
\end{multline}%
Finally, posing $\eta =S\tilde{\eta}S^{\dagger }=SV\Theta _{\mathfrak{F}%
}S^{\dagger }\Omega $ one obtains

\begin{equation*}
\eta H\eta ^{-1}=SV\Theta _{\mathfrak{F}}S^{\dagger }\Omega H\left( SV\Theta
_{\mathfrak{F}}S^{\dagger }\Omega \right) ^{-1}=S\tilde{H}^{\dagger
}S^{-1}=H^{\dagger }.
\end{equation*}

$ii)\Rightarrow iii)$. If $H$ is (weakly) pseudo-Hermitian, the eigenvalues
of $H$ are either real or come in complex-conjugate pairs and the geometric
multiplicity and the Jordan dimensions of the complex-conjugate eigenvalues
coincide (see the remark below Theorem 1). Then, one can easily see,
recalling the definition of the operator $U$ provided in the proof of
Theorem 1 (Eq.(11)) and Eqs. (10) and (17), that

\begin{equation*}
\Theta _{\mathfrak{F}}\tilde{H}\Theta _{\mathfrak{F}}=U\tilde{H}U.
\end{equation*}

Hence the antilinear operator

\begin{eqnarray*}
\tilde{\Omega} &=&\Theta _{\mathfrak{F}}U=U\Theta _{\mathfrak{F}%
}=\sum_{n_{0}}\sum_{a=1}^{d_{n_{0}}}\sum_{i=1}^{p_{n_{0},a}}|u_{n_{0}},a,i%
\rangle K\left\langle u_{n_{0}},a,i\right| + \\
&&\sum_{n_{+},n_{-}}\sum_{a=1}^{d_{n_{+}}}\sum_{i=1}^{p_{n_{+},a}}(\left|
u_{n_{+}},a,i\right\rangle K\left\langle u_{n_{-}},a,i\right| +\left|
u_{n_{-}},a,i\right\rangle K\left\langle u_{n_{+}},a,i\right| )
\end{eqnarray*}%
commutes with $\tilde{H}$. Moreover, $\tilde{\Omega}$ is involutory, as one
can immediately verify by using the explicit expression of $\tilde{\Omega}$
in the previous equation. Then, it follows immediately (recalling Eq. (18)
and observing that $\Theta _{\mathfrak{F}}S^{\dagger }=S^{\dagger }\Theta _{%
\mathfrak{E}}$ ) that 
\begin{equation}
\hat{\Omega}=S^{\dagger -1}\tilde{\Omega}S^{\dagger }=S^{\dagger -1}U\Theta
_{\mathfrak{F}}S^{\dagger }=S^{\dagger -1}US^{\dagger }\Theta _{\mathfrak{E}}
\end{equation}%
commutes with $H$ and is involutory.

$iii)\Rightarrow iv)$. (See Prop.5 in \cite{so}, where an analogous
statement has been proven, referring to diagonalizable operators).

If we denote by $L$ the linear part of $\widehat{\Omega }$, \textit{i.e.}, $%
\widehat{\Omega }=LK$ (where $K$ is the complex conjugation operator), then $%
\widehat{\Omega }^{2}=\mathbf{1}$ implies $LL^{\ast }=\mathbf{1}$ and this
is possible if and only if an $M$ exists such that $L=MM^{\ast -1}$\cite{asc}%
. Then $[H,\widehat{\Omega }]=0$ implies $HMM^{\ast -1}=MM^{\ast -1}H^{\ast
} $, hence

\begin{equation*}
M^{-1}HM=(M^{\ast -1}H^{\ast }M^{\ast })=(M^{-1}HM)^{\ast }.
\end{equation*}

$iv)\Rightarrow i)$. Trivially, every operator which assumes a real form in
some basis $\mathfrak{B}$ commutes with the conjugation associated with $%
\mathfrak{B}$.$\blacksquare $

\bigskip

\textbf{Remark}. Note that the equivalence $i)\Longleftrightarrow iv)$ we
proven above clearly restates precisely a similar (seemingly, more general)
result in literature, according to which \textit{whenever }$H$\textit{\
commutes with an antiunitary symmetry }$A$\textit{\ such that }$A^{2k}=1$%
\textit{\ }$\left( k\text{ odd}\right) $\textit{, it is possible to
construct a basis in which the matrix elements of }$H$\textit{\ are real}. %
\cite{ben0}

\bigskip

\section{The Kramers degeneracy}

On the basis of the above-stated theorem (in particular, by the implication $%
i)\Rightarrow ii)$\ ) one can conclude that any time-reversal invariant
(diagonalizable or not) Hamiltonian $H$ must belong to the class of
pseudo-Hermitian Hamiltonians. The converse does not hold in general, since
not always one can interpret the antilinear symmetry $\Omega $ of $H$ as the
time-reversal operator $T$ ; furthermore, it is well known that in case of
fermionic systems%
\begin{equation*}
T^{2}=-\mathbf{1}
\end{equation*}%
and the above theorem, whereas it assures the existence of an involutory
antilinear symmetry, does not say anything about the existence of a symmetry
like $T$.

In order to go more deeply into the matter, we can now state the following

\textbf{Theorem 4.}\textit{\ Let }$H$\textit{\ be a linear operator with a
discrete spectrum. Then, the following conditions are equivalent:}

\textit{i) an antilinear operator }$\mathfrak{T}$\textit{\ exists such that }%
$\left[ H,\mathfrak{T}\right] =0$\textit{\ , with }$\mathfrak{T}^{2}=-1;$

\textit{ii) }$H$\textit{\ is pseudo-Hermitian and the Jordan blocks
associated with any real eigenvalue occur in pair [i.e., for any couple }$%
E_{n_{0}},a,$ \textit{the number} $k(n_{0},a)$\textit{\ is even} \textit{\
(see Sec. 2)}].

\textbf{Proof}. Let us assume that condition $i)$ holds; then, by Theorem 3, 
$H$ is pseudo-Hermitian, hence its eigenvalues are either real or come in
complex-conjugate pairs and the geometric multiplicity and the Jordan
dimensions of the complex-conjugate eigenvalues coincide (see Theorem 2).

Let now $\left| \psi _{n_{0}},a,1\right\rangle $ be an eigenvector of $H$;
then, $\mathfrak{T}\left| \psi _{n_{0}},a,1\right\rangle $ too is an
eigenvector of $H$, corresponding to the same eigenvalue $E_{n_{0}}$, and
linearly independent from $\left| \psi _{n_{0}},a,1\right\rangle $ .
(Indeed, assume that $\mathfrak{T}$ $\left| \psi _{n_{0}},a,1\right\rangle
=\alpha \left| \psi _{n_{0}},a,1\right\rangle $ for some $\alpha \in \mathbf{%
C};$ applying $\mathfrak{T}$ one gets $\left| \psi _{n_{0}},a,1\right\rangle
=-\left| \alpha \right| ^{2}\left| \psi _{n_{0}},a,1\right\rangle $, which
is impossible.)

If $\left| \psi _{n_{0}},b,1\right\rangle $ is another eigenvector of $H$,
linearly independent from $\left| \psi _{n_{0}},a,1\right\rangle $ and $%
\mathfrak{T}\left| \psi _{n_{0}},a,1\right\rangle $, also $\mathfrak{T}%
\left| \psi _{n_{0}},b,1\right\rangle $ is linearly independent from all
three; otherwise, applying once again $\mathfrak{T}$ to the relation 
\begin{equation*}
\alpha \left| \psi _{n_{0}},a,1\right\rangle +\beta \mathfrak{T}\left| \psi
_{n_{0}},a,1\right\rangle +\gamma \left| \psi _{n_{0}},b,1\right\rangle
+\delta \mathfrak{T}\left| \psi _{n_{0}},b,1\right\rangle =0
\end{equation*}%
we could eliminate, for instance, $\mathfrak{T}\left| \psi
_{n_{0}},b,1\right\rangle $, thus obtaining a linear dependence between $%
\left| \psi _{n_{0}},a,1\right\rangle ,\mathfrak{T}\left| \psi
_{n_{0}},a,1\right\rangle $ and $\left| \psi _{n_{0}},b,1\right\rangle $,
contrary to the previous hypothesis.

We can conclude, iterating this procedure, that the geometric multiplicity $%
d_{n_{0\text{ }}}$of $E_{n_{0}}$ must be necessarily even. Moreover, one can
always assume that, for a suitable choice of the basis vectors, $\mathfrak{T}%
\left| \psi _{n_{0}},a,1\right\rangle \equiv \left| \psi _{n_{0}},a^{\prime
},1\right\rangle $ for some $a^{\prime }$.

Let us consider now the subset of vectors $\{\left| \psi
_{n_{0}},a,i\right\rangle ,i=1,...p_{n_{0},a}\}$. They constitute a basis in
the subspace associated with the Jordan block $J_{a}\left( E_{n_{0}}\right) $%
; then by hypothesis one has

\begin{equation*}
\sum_{i=1}^{p_{n_{0},a}}\alpha _{i}\left| \psi _{n_{0}},a,i\right\rangle =0%
\text{ \ }\Leftrightarrow \text{ \ }\alpha _{i}=0\text{ }\forall
i=1,...p_{n_{0},a}.
\end{equation*}%
Applying $\mathfrak{T}$\ to the previous equation, one obtains

\begin{equation*}
\sum_{i=1}^{p_{n_{0},a}}\alpha _{i}^{\ast }\mathfrak{T}\left| \psi
_{n_{0}},a,i\right\rangle =0\text{ \ }\Leftrightarrow \text{ \ }\alpha _{i}=0%
\text{ }\forall i=1,...p_{n_{0},a},
\end{equation*}%
hence, the vectors $\left\{ \mathfrak{T}\left| \psi
_{n_{0}},a,i\right\rangle \equiv \left| \psi _{n_{0}},a^{\prime
},i\right\rangle ,i=1,...p_{n_{0},a}\right\} $ too are linearly independent,
and $p_{n_{0},a}=\dim J_{a}\left( E_{n_{0}}\right) \leq p_{n_{0},a^{\prime
}}=\dim J_{a^{\prime }}\left( E_{n_{0}}\right) .$On the other hand, applying 
$\mathfrak{T}$ to the basis vectors $\left\{ \mathfrak{T}\left| \psi
_{n_{0}},a,i\right\rangle \right\} $ of the subspace associated with $%
J_{a^{\prime }}\left( n_{0}\right) $, one obtains that the dimensions of the
two blocks must coincide, hence $J_{a}\left( n_{0}\right) $ and $%
J_{a^{\prime }}\left( n_{0}\right) $ are identical.

(Alternatively, the same result can be obtained by applying $\mathfrak{T}$
to both members of Eq. (4)).

Conversely, let condition $ii)$ hold; then $H$ assumes the form

\begin{eqnarray*}
H
&=&\sum_{n_{0}}\sum_{a=1}^{d_{n_{0}}/2}[E_{n_{0}}\sum_{i=1}^{p_{n_{0},a}}(%
\left| \psi _{n_{0}},a,i\right\rangle \left\langle \phi _{n_{0}},a,i\right|
+\left| \psi _{n_{0}},a+d_{n_{0},a}/2,i\right\rangle \left\langle \phi
_{n_{0}},a+d_{n_{0},a}/2,i\right| ) \\
&&\sum_{i=1}^{p_{n_{0},a-1}}(\left| \psi _{n_{0}},a,i\right\rangle
\left\langle \phi _{n_{0}},a,i+1\right| +\left| \psi
_{n_{0}},a+d_{n_{0},a}/2,i\right\rangle \left\langle \phi
_{n_{0}},a+d_{n_{0},a}/2,i+1\right| )]+ \\
&&\sum_{n_{+},n_{-}}\sum_{a=1}^{d_{n+}}[\sum_{i=1}^{p_{n_{+},a}}(E_{n_{+}}%
\left| \psi _{n_{+}},a,i\right\rangle \left\langle \phi _{n+},a,i\right|
+E_{n_{+}}^{\ast }\left| \psi _{n_{-}},a,i\right\rangle \left\langle \phi
_{n_{-}},a,i\right| )+ \\
&&\sum_{i=1}^{p_{n_{+},a-1}}(\left| \psi _{n_{+}},a,i\right\rangle
\left\langle \phi _{n+},a,i+1\right| +\left| \psi _{n_{-}},a,i\right\rangle
\left\langle \phi _{n_{-}},a,i+1\right| )].
\end{eqnarray*}

Let us denote by $\mathfrak{T}$ the following antilinear operator: 
\begin{eqnarray}
\mathfrak{T} &=&\sum_{n_{0}}\sum_{a=1}^{d_{n_{0}}/2}\sum_{i=1}^{p_{n_{0},a}}%
\left( \left| \psi _{n_{0}},a,i\right\rangle K\left\langle \phi
_{n_{0}},a+d_{n_{0},a}/2,i\right| -\left| \psi
_{n_{0}},a+d_{n_{0},a}/2,i\right\rangle K\left\langle \phi
_{n_{0}},a,i\right| \right) +  \notag \\
&&\sum_{n_{+},n_{-}}\sum_{a=1}^{d_{n+}}[\sum_{i=1}^{p_{n_{+},a}}\left(
\left| \psi _{n_{-}},a,i\right\rangle K\left\langle \phi _{n+},a,i\right|
-\left| \psi _{n_{+}},a,i\right\rangle K\left\langle \phi
_{n_{-}},a,i\right| \right) ,
\end{eqnarray}%
where the operator $K$ acts transforming each complex number on the right
into its complex-conjugate. Then, one easily obtains, by inspection, that $%
\left[ H,\mathfrak{T}\right] =0$ and $\mathfrak{T}^{2}=-\mathbf{1}.$ $%
\blacksquare $

\bigskip

Recalling that the algebraic multiplicity of any $E_{n}$ is $%
g_{n}=\sum_{a=1}^{d_{n}}p_{n,a}$ , from Theorem 4 in particular it follows
that whenever\textit{\ \ a pseudo-Hermitian operator }$H$\textit{\ admits an
antilinear symmetry }$\mathfrak{T}$\textit{\ with }$\mathfrak{T}^{2}=-1$%
\textit{, both the geometric and the algebraic multiplicity of any real
eigenvalue of }$H$\textit{\ is even.}

The above-mentioned theorem generalizes an analogous theorem stated from the
authors (and referring to diagonalizable pseudo-Hermitian operators)\cite{kr}%
, which in turn generalizes from various point of view the Kramers theorem
on the degeneracy of any fermionic (Hermitian) Hamiltonian. Hence, by an
abuse of language, we will continue to denote as ''\textit{Kramers degeneracy%
}'' this typical feature of real eigenvalues of pseudo-Hermitian operators
admitting a symmetry like $\mathfrak{T.}$

\section{Concluding remarks}

Basing on Theorem 4, we can quickly test the $T$-invariance properties of
pseudo-Hermitian Hamiltonians. Indeed, let us consider for instance the
operator%
\begin{equation*}
H_{eff}=\left( 
\begin{array}{cc}
E & ir \\ 
is & E%
\end{array}%
\right) (E,r,s\in \mathbf{R})
\end{equation*}%
which we already discussed elsewhere \cite{kr}, and which arises in the
modified Mashhoon model \cite{r}, where one introduces a ($T$-violating)
spin-rotation coupling to explain the muon's anomalous $g$ factor.

This Hamiltonian (as long as it is diagonalizable) is time-reversal
violating \cite{kr}; however, for some choice of parameter values (for
instance, $r\neq s=0$), $H_{eff}$ is no longer diagonalizable. Now, on the
basis of Theorem 4 we can conclude that also for such values $H_{eff}$
cannot admit an antilinear symmetry $\mathfrak{T}$ such that $\mathfrak{T}%
^{2}=-\mathbf{1}$ (hence, $H_{eff}$ \ cannot be $T$-invariant). In fact,
being the geometric multiplicity of its eigenvalue $E$ odd, condition $ii)$
of Theorem 4 does not hold. We recall however that we obtained the same
result by a straightforward calculation \cite{kr}.

Finally, we note that in a symmetry-adapted basis $\left\{ \left| \psi
_{n}\right\rangle ,\mathfrak{T}\left| \psi _{n}\right\rangle \right\} $ the
matrix of any pseudo-Hermitian operator $H$ , satisfying condition $ii)$ of
Theorem 4, assumes a symplectic form. This property, in the Hermitian case,
is often used in order to simplify some electronic-structure calculations
occurring for instance in molecular or solid-state physics. \cite{es}

\end{document}